\documentclass[11pt,dvips]{article}

\usepackage{graphicx}

\newcommand{\eV}{\mathrm{e\kern -0.1em V}}
\newcommand{\MeV}{\mathrm{Me\kern -0.1em V}}
\newcommand{\GeV}{\mathrm{Ge\kern -0.1em V}}
\newcommand{\TeV}{\mathrm{Te\kern -0.1em V}}

\begin{document}

\title{{\bf Colour Reconnection in W Decays}}

\author{{\it Paul de Jong} \\ \\
NIKHEF, P.O.Box 41882, 1009 DB Amsterdam, the Netherlands \\
E-mail: Paul.de.Jong@cern.ch \\ \\
{\it Contribution to the Proceedings of the XXX International} \\
{\it Conference on High Energy Physics ICHEP2000, Osaka, Japan}}

\date{}

\maketitle

\abstract{
The studies of colour reconnection in $e^+ e^- \rightarrow 
\mathrm{W^+ W^-} \rightarrow q \bar{q}' q \bar{q}'$ events at LEP are 
reviewed. It is shown that the analysis of the
particle- and energy flow between jets is sensitive to
realistic model predictions. The effects on the W mass measurement
are discussed. Most results are preliminary.}

\section{Colour Reconnection}

With some 450 pb$^{-1}$ per experiment
already recorded at $\sqrt{s} = 183-202$ $\GeV$, and more to come at higher
energies, each of the four LEP experiments have selected up to now some 3200
WW $\rightarrow \, qqqq$ and some 2500 WW $\rightarrow \, qq \ell \nu$
candidate events. 
The mass and width of the W boson are measured from the
kinematics of W decay products.
Any energy-momentum exchange between
W decay products not well simulated in Monte Carlo will affect the
W mass and width measurement. Conventional MC's treat the two
$q \bar{q}'$ systems in a WW $\rightarrow \, q \bar{q}' q \bar{q}'$ event 
as independent. However, QCD interconnections, or colour
reconnection (CR) can be expected~\cite{cr}.
Perturbative CR effects are estimated to be small~\cite{sk};
CR is a non-perturbative hadronization uncertainty that can only
be studied in the context of various models.

CR models being used in these studies are those implemented in
PYTHIA, ARIADNE and HERWIG. The models in PYTHIA, SK I, SK II and
SK II', are based on reconfiguration of overlapping or crossing
strings~\cite{sk}. In the SK I model, the probability of reconnection
is calculated as $P_{\mathrm{reco}} = 1 - e^{(-k_i O)}$, where $O$
is the overlap of two finite strings, and $k_i$ is a free parameter.
In the SK II and II' models, the string has no lateral dimension, and
strings are reconnected when they cross.
The ARIADNE models are based on
rearrangement of colour dipoles if this reduces the string 
length~\cite{ar}. 
It should be noted that these models also affect \mbox{LEP 1} data and
are disfavoured from an OPAL study of the properties of quark- and gluon
jets~\cite{oa}. 

A reconfiguration of the colour flow is expected to change the 
(charged) particle multiplicity (typically decreasing it by
0.2 to 0.9 units), especially at low momentum, and more
specifically between jets associated to the same W.

\section{Multiplicities}

\subsection{Inclusive Charged Multiplicity}

The charged multiplicity in WW events is measured by all four LEP
experiments from charged tracks in the tracking 
system~\cite{a1,d1,l1,o1}. The
track multiplicity distribution is corrected to a charged particle
multiplicity distribution by a matrix unfolding procedure. Alternatively,
the multiplicity as a function of momentum (fragmentation function) or
$p_T$ is determined and corrected bin-by-bin. The results are shown
in Table~\ref{tab:multi}.

\begin{table}[tb]
\begin{center}
\begin{tabular}{|l|l|l|l|} \hline
    & $<N_{ch}^{4q}>$ & $<N_{ch}^{2q}>$ & $\Delta <N_{ch}>$ \\ \hline 
OPAL 183 $\GeV$ & $39.4 \pm 0.5 \pm 0.6$ & $19.3 \pm 0.3 \pm 0.3$ & $+0.7 \pm 0.8 \pm 0.6$ \\ 
OPAL 189 $\GeV$ & $38.31 \pm 0.24 \pm 0.37$ & $19.23 \pm 0.19 \pm 0.19$ & $-0.15 \pm 0.44 \pm 0.38$ \\ \hline 
L3 183-202 $\GeV$ & $37.90 \pm 0.14 \pm 0.41$ & $19.09 \pm 0.11 \pm 0.21$ & $-0.29 \pm 0.26 \pm 0.30$ \\ \hline 
DELPHI 183 $\GeV$ & $38.11 \pm 0.57 \pm 0.44$ & $19.78 \pm 0.49 \pm 0.43$ &  \\ 
DELPHI 189 $\GeV$ & $39.12 \pm 0.33 \pm 0.36$ & $19.49 \pm 0.31 \pm 0.27$ &  \\ \hline 
ALEPH 183-202 $\GeV$ & $35.75 \pm 0.13 \pm 0.52$ & $17.41 \pm 0.12 \pm 0.15$ & $+0.93 \pm 0.27 \pm 0.29$ \\ \hline
\end{tabular}
\end{center}
\caption{{\small Average charged multiplicity in $qqqq$ events, 
$<N_{ch}^{4q}>$,
in $qq \ell \nu$ events, $<N_{ch}^{2q}>$, and the difference
$\Delta <N_{ch}> = <N_{ch}^{4q}> - 2 <N_{ch}^{2q}>$, as measured by
the four LEP experiments. The ALEPH results are quoted within detector
acceptance and not corrected to full phase space;
DELPHI prefers to quote the ratio $R = <N_{ch}^{4q}>/
2<N_{ch}^{2q}>$, see text.}}
\label{tab:multi}
\end{table}

The difference $\Delta <N_{ch}> = <N_{ch}^{4q}> - 2 <N_{ch}^{2q}>$ is
also given in Table~\ref{tab:multi}. DELPHI prefers to quote the
ratio $R = <N_{ch}^{4q}>/2<N_{ch}^{2q}> = 0.990 \pm 0.015 \pm 0.011$.
Combining the results, it can be concluded that $\Delta <N_{ch}>$ is
consistent with 0 within an error of 0.3-0.4. A proper average is
difficult due to differences in the definition, and the correlated
systematics; the size of these correlated systematics (0.2-0.3), which
is of the same size as the CR effects, limits the sensitivity of this
method.

\subsection{Fragmentation Function}

By studying the particle multiplicity as a function of $x_p = 2 p /\sqrt{s}$,
or $\xi = - \log (x_p)$, one can study the low momentum region $p < 1$ $\GeV$
where CR effects predominantly reside, but at the cost of reduced statistics.
No significant effects at low $x_p$ are observed by any of the experiments.

\subsection{Heavy Hadrons}
 
Massive particles, like $K^{\pm}$ or (anti)protons, are more sensitive
to CR effects than pions, by a factor 2 to 3.
However, this is counterbalanced by the decreased statistics.
DELPHI~\cite{d1} and OPAL~\cite{o2} have studied the production of 
heavy hadrons in $qqqq$ and $qq \ell \nu$ events and observe no significant 
differences.

\section{Particle Flow}
 
A more promising technique to study CR appears to be the study of
the particle- or energy flow between jets from the same W and
between different W's, in analogy to studies of the string effect~\cite{dd}.

\begin{figure}[tb]
\begin{center}
\includegraphics[width=80mm]{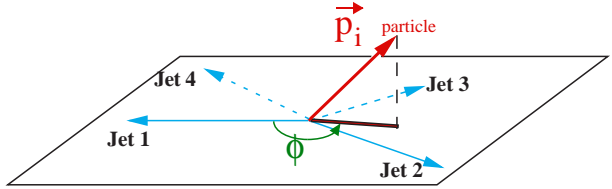}
\end{center}
\caption{Construction of the particle flow.}
\label{fig:flow}
\end{figure}

The construction of the particle flow is explained in Figure~\ref{fig:flow}.
A jet-finder is used to construct four jets in $qqqq$ events. Each pair of
jets defines a plane onto which all reconstructed particles in the event
are projected; for the energy flow
weighted with the particle energy. In the preliminary studies submitted to this
conference, L3~\cite{l1} and ALEPH~\cite{a2} use strong cuts on the 
angles between jets to
select topologies with well separated jets and planar-like events, and
obtain a selection efficiency of $\sim 15$\%;
OPAL~\cite{o3} uses less restrictive cuts and a jet-pairing likelihood
that gives a higher efficiency of $\sim 42$\%, but selects also topologies
with less clear separation between CR models.
The flow is symmetrized with respect to the choice of jet-pairs, and
particle angles between jets are rescaled between 0 and 1.

CR models indeed show a depletion of the particle flow between jets from the
same W, and an increase between jets from different W's, as 
naively expected. It is convenient to average the flow in the two regions 
between jets from the same W (regions j1-j2 and j3-j4), and to do the same for 
the flow in the two regions between jets from different W's (regions j2-j3 and
j4-j1). Subsequently,
the ratio of these within-W/between-W flows is taken
as a function of the rescaled particle angle.

\begin{figure}[tb]
\begin{center}
\includegraphics[width=100mm]{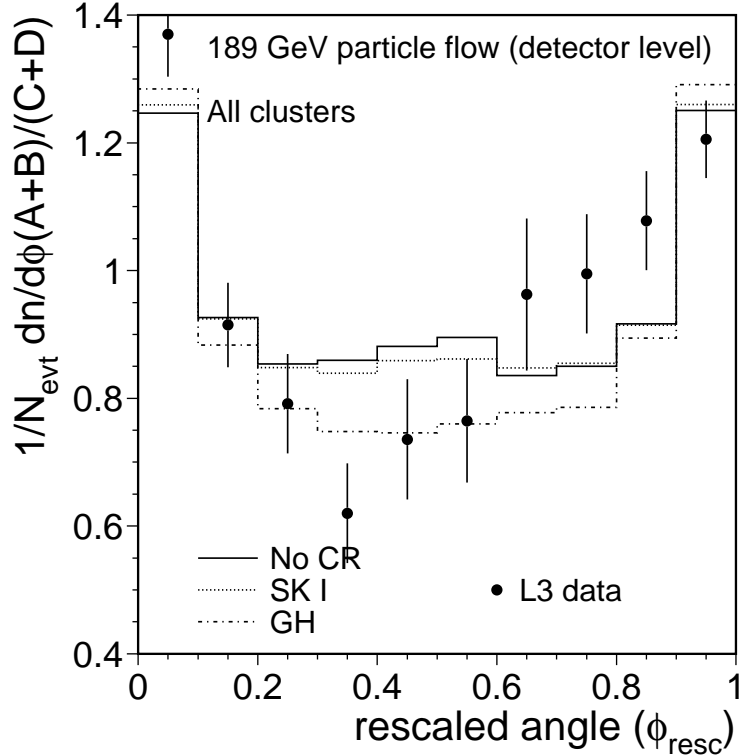}
\end{center}
\caption{{\small Ratio of the particle flows between jets from the same W and
between jets from different W's, as a function of the rescaled angle,
for L3 data at $\sqrt{s} = 189$ $\GeV$ and Monte Carlo.}}
\label{fig:l3flow}
\end{figure}

L3 has studied 176 pb$^{-1}$ of data taken at $\sqrt{s} = 189$ $\GeV$.
The ratio of the particle flow between jets
from the same W and between jets from different W's is shown in
Figure~\ref{fig:l3flow}, for data, PYTHIA without CR, and PYTHIA
with SK I and GH, as a function of the rescaled angle 
$\phi_{\mathrm{resc}}$. With this data sample only, a
sensitivity of 3.5 $\sigma$ to SK I ($k_i = 1000$) and 
1.0 $\sigma$ to SK I ($k_i = 0.6$) is reached. Varying the fraction
of reconnected events, and calculating the $\chi^2$ for the data-MC
comparison, a fraction of $\sim$40\% of reconnected events in the
SK I model is favoured, and the No-CR scenario is
disfavoured at 1.7~$\sigma$.

\begin{figure}[tb]
\begin{center}
\includegraphics[width=100mm]{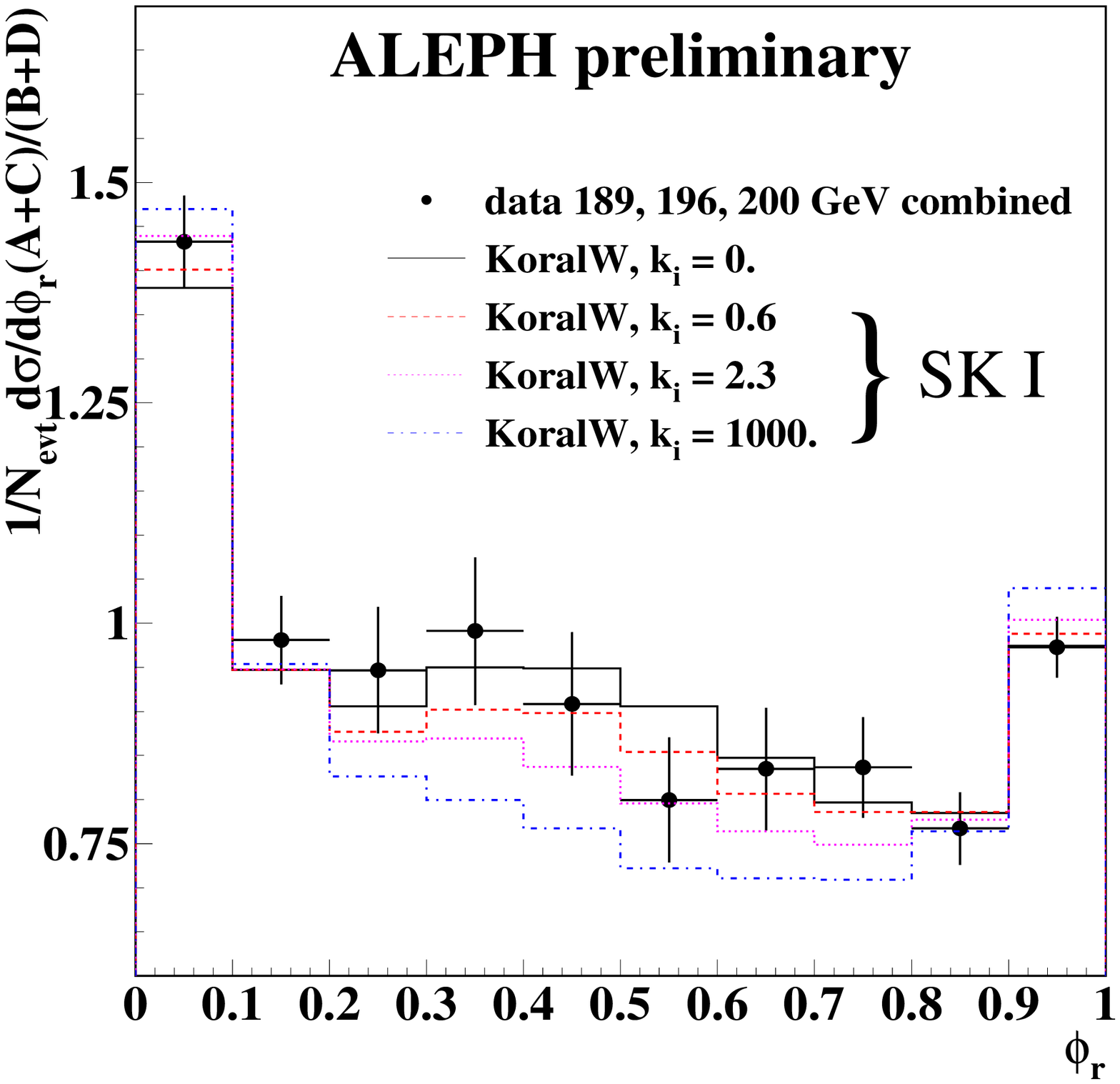}
\end{center}
\caption{{\small Ratio of the particle flows between jets from the same W and
between jets from different W's, as a function of the rescaled angle,
for ALEPH data at $\sqrt{s} = 189-200$ $\GeV$ and Monte Carlo.}}
\label{fig:alephflow}
\end{figure}

ALEPH has analyzed 347 pb$^{-1}$ of data taken at $\sqrt{s} = 189-200$ $\GeV$.
Their particle flow ratio is shown in
Figure~\ref{fig:alephflow}, for data, KORALW without CR, and KORALW
with the SK I model for various values of $k_i$.
Varying $k_i$, ALEPH finds the best data-MC agreement
for $k_i \approx 0.25$, and puts a 1 $\sigma$ upper limit on
$k_i$ of $1.4$ which corresponds to 45\% of reconnected events at
$\sqrt{s} = 189$ $\GeV$.

OPAL has studied 183 pb$^{-1}$ of data taken at $\sqrt{s} = 189$ $\GeV$,
and find sensitivities of 4.0 $\sigma$
for SK I ($k_i = 100$), 1.1 $\sigma$ for SK I ($k_i = 0.9$),
0.4~$\sigma$ for SK II and II', and 0.5-1.8~$\sigma$ for AR2 and AR3.
As a cross-check, OPAL uses the strong cuts like
L3 and ALEPH, and observes slightly smaller sensitivities. The actual
data is ambiguous, and prefers some reconnection in the default analysis,
but no CR in the cross-check analysis. This, and in particular the
role of the background, will be further studied.

DELPHI is also working on a similar analysis, but was not yet able to
submit results to this conference.

With the full LEP 2 data sample, and combining all experiments, a
further gain in sensitivity by a factor $\sim 3.5$ can be expected.

\section{Effect on $M_{\mathrm{W}}^{qqqq}$}

The estimates for $\Delta M_{\mathrm{W}}^{qqqq}$ from the individual
experiments calculated with their own Monte Carlo samples are
summarized in Table~\ref{tab:deltamw}~\cite{mw}. 
A difference in reconnection probability
in the SK II model can be expected from differences in the parton shower
cutoff scale $Q_0$~\cite{sk}, which ranges between 1.0 $\GeV$ (L3) and 1.9
$\GeV$ (OPAL). 

\begin{table}[tb]
\begin{center}
\begin{tabular}{|c|c|c|c|c|}
\hline
\hspace{5mm} & OPAL & L3 & DELPHI & ALEPH \\
\hline 
SK I & $+66 \pm 8$ ($35.1$\%) & $+29 \pm 15$ ($32.1$\%) &  $+46 \pm 2$ ($35.9$\%)& $+30 \pm 10$ ($29.2$\%) \\
          & ($k_i = 0.9$) & ($k_i = 0.6$) & ($k_i = 0.65$) & ($k_i = 0.65$) \\
SK II & $+3 \pm 8$ ($19.8$\%) & $-5 \pm 15$ ($32.4$\%) & $-2 \pm 5$ & $+6 \pm 8$ ($29.2$\%) \\
SK II' & $+10 \pm 8$ ($17.6$\%) & $-33 \pm 15$ ($28.8$\%) & \hspace{5mm} & $+4 \pm 8$ ($26.7$\%) \\
AR 2 & $+85 \pm 8$ ($50.3$\%) & $+106 \pm 26$ & $+28 \pm 6$ & $+21 \pm 19$ \\
AR 3 & $+140 \pm 10$ ($62.3$\%) & $+170 \pm 26$ & $+55 \pm 6$ & $+34 \pm 34$ \\
HERWIG & \hspace{5mm} & \hspace{5mm} & \hspace{5mm} & $+20 \pm 10$ \\
\hline
\end{tabular}
\end{center}
\caption{{\small Experimental estimates of $\Delta M_{\mathrm{W}}^{qqqq}$,
in $\MeV$, from
the four LEP experiments, as calculated with their own implementations
of various CR models at $\sqrt{s} = 189$ $\GeV$. The fraction of
reconnected events in each sample is given between brackets.}}
\label{tab:deltamw}
\end{table}

Common samples of KORALW + JETSET Monte Carlo events, with (SK I)
and without colour reconnection, have been generated for the four experiments; 
each experiment then passed these samples though detector simulation, event
selection and analysis procedures. The resulting mass shifts found by
the experiments were equal within errors, 
and a correlation between experiments of close to
100\% was found. In view of this,
further LEP collaboration will be needed to fully understand the 
differences in Table~\ref{tab:deltamw}, especially in
the ARIADNE estimates. 
For the LEP $M_{\mathrm{W}}$ combination, a CR systematic error of 50 $\MeV$
was used, fully correlated between experiments.
 
Estimates of the CR effect on the W width in the $qqqq$ channel
range between +40 and +70 $\MeV$ in the SK models~\cite{mw}.

The studies of the particle flow between jets have proven
to be sensitive to realistic CR model predictions.
These studies will thus directly measure the amount of CR in the data.
In models with a free parameter, such as SK I, this parameter
can be measured from data; a calibration curve of mass shift
versus $k_i$ can be used to estimate the CR systematic error. 
Already ALEPH, with a
1 $\sigma$ upper limit on $k_i$ of 1.4, puts a 1 $\sigma$ upper limit
on $\Delta M_{\mathrm{W}}^{qqqq}$ in the SK~I model of 40 $\MeV$~\cite{a2}.
With the full LEP 2 data sample and combining all experiments,
the CR systematic error on $M_{\mathrm{W}}$ is likely to be
below that, and, most important, actually measured from data.

\end{document}